# Translational and reorientational dynamics in deep eutectic solvents

D. Reuter,[1] P. Münzner,[2] C. Gainaru,[2] P. Lunkenheimer,[1,a)] A. Loidl,[1] and R. Böhmer[2]

**AFFILIATIONS**

[1]Experimental Physics V, Center for Electronic Correlations and Magnetism, University of Augsburg, 86135 Augsburg, Germany

[2]Fakultät Physik, Technische Universität Dortmund, 44221 Dortmund, Germany

[a)]**Author to whom correspondence should be addressed:** peter.lunkenheimer@physik.uni-augsburg.de

**ABSTRACT**

We performed rheological measurements of the typical deep eutectic solvents (DESs) glyceline, ethaline, and reline in a very broad temperature and dynamic range, extending from the low-viscosity to the high-viscosity supercooled-liquid regime. We find that the mechanical compliance spectra can be well described by the random free-energy barrier hopping model, while the dielectric spectra on the same materials involve significant contributions arising from reorientational dynamics. The temperature-dependent viscosity and structural relaxation time, revealing non-Arrhenius behavior typical for glassy freezing, are compared to the ionic dc conductivity and relaxation times determined by broadband dielectric spectroscopy. For glyceline and ethaline we find essentially identical temperature dependences for all dynamic quantities. These findings point to a close coupling of the ionic and molecular translational and reorientational motions in these systems. However, for reline the ionic charge transport appears decoupled from the structural and reorientational dynamics, following a fractional Walden rule. Especially, at low temperatures the ionic conductivity in this DES is enhanced by about one decade compared to expectations based on the temperature dependence of the viscosity. The results for all three DESs can be understood without invoking a revolving-door mechanism previously considered as a possible charge-transport mechanism in DESs.

## I. INTRODUCTION

Deep eutectic solvents (DES) are promising alternatives to ionic liquids in a variety of applications, e.g., as "green solvents" in material synthesis or as electrolytes in electrochemical devices like batteries or solar cells.[1,2,3,4,5,6,7,8,9,10,11] Many of them can be produced at low cost and are superior concerning environmental friendliness and renewability. DESs are mixtures of two or more components leading to the typical melting-point reduction of eutectics which makes them liquid at room temperature. Typical for eutectic mixtures, they usually can be easily supercooled and exhibit a glass transition at low temperatures,[4,12,13] an often neglected fact that also can affect their room-temperature properties.[13] The most common types of DESs are composed of a molecular hydrogen-bond donor (HBD) and a hydrogen-bond acceptor, often a quaternary ammonium salt. For electrochemical applications, their ionic dc conductivity is an important benchmark quantity ($\gtrsim 10^{-4}\ \Omega^{-1}\ cm^{-1}$ at room temperature). Thus, understanding the ionic charge-transport mechanism in DESs and finding ways for its optimization are crucial tasks of current research on these materials.

Notably, in addition to the translational dynamics of the ionic charge carriers, in most DESs also reorientational dynamics should exist because the HBDs (e.g., glycerol or urea) and the ion species of the added salt (e.g., the cation of the frequently-used choline chloride) often are asymmetric molecules with rotational degrees of freedom. This type of dynamics can be investigated by dielectric spectroscopy because asymmetric molecules and ions usually possess dipole moments, whose reorientations lead to typical features in spectra of the complex dielectric permittivity.[14,15] Moreover, the simultaneous detection of the ionic conductivity by this method allows for a direct comparison of both dynamics. Nevertheless, only few dielectric investigations of DESs were reported until now.[12,13,16,17,18,19,20] Interestingly, a recent dielectric study of three DESs, performed by some of the present authors,[13] revealed a close correlation of their reorientational dynamics with the ionic dc conductivity for two systems (glyceline and ethaline) while some minor but significant deviations showed up for the third material (reline). Moreover, the temperature dependence of both the rotational and ionic dynamics in these systems were found to exhibit the characteristics of glassy freezing. The detected rotation-translation coupling could indicate that the often-neglected molecular rotation modes in these materials are relevant not only from an academic but also from an application point of view as they may affect the ionic dc conductivity.

However, by dielectric spectroscopy alone, it is not possible to conclude about the microscopic origin of this coupling. Is this of an indirect nature, arising from a simultaneous coupling of both dynamics to the viscosity? Or does it occur directly, via a "revolving-door" mechanism, as considered for plastic-crystal electrolytes[21,22,23] and also proposed for ionic liquids?[24] In the latter scenario, the reorientation of the asymmetric molecules within the material is assumed to open up transient gaps which are expected to enhance the ionic mobility, and, thus, the conductivity, compared to the value expected for a particle diffusing through a viscous medium. On the other hand, for the purely viscosity-driven scenario, it is intuitively clear that the translational motion of particles through a viscous medium should be related to its viscosity, just as the rotational motion of asymmetric molecules within this medium. In the simplest case, both dynamics, which can be quantified by the reorientational relaxation time $\tau_\varepsilon$ and the dc conductivity $\sigma_{dc}$, should then be related to the viscosity via $\eta \propto \tau_\varepsilon \propto 1/\sigma_{dc}$. More precisely, one can combine the Stokes-Einstein relation (predicting $D_t \propto T/\eta$, with $D_t$ the translational diffusion coefficient), the Debye-Stokes-Einstein relation ($1/\tau_\varepsilon \propto D_r \propto T/\eta$ with $D_r$ denoting the rotational diffusion coefficient), and the Nernst-Einstein relation ($\sigma_{dc} \propto D_i/T$ with $D_i$ the ionic diffusion coefficient),[25,26,27,28] to arrive at a similar but slightly different relation: $\eta \propto T\tau_\varepsilon \propto 1/\sigma_{dc}$ (assuming $D_i = D_t$). However, the additional temperature factor applied to $\tau_\varepsilon$ can essentially be neglected: Even for extensive studies of $\tau_\varepsilon(T)$, the covered temperatures vary by less than a factor of two while the relaxation time changes by many decades. One should note, however, that for various liquids, especially so-called fragile glass formers, deviations from the Stokes-Einstein, Debye-Stokes-Einstein, and Nernst-Einstein relations were reported.[26,28,29,30,31] Different explanations were proposed, e.g., in terms of the heterogeneity of glassy dynamics[26,29,30] or within other models.[32,33] Correspondingly, translation-rotation coupling or decoupling also are long-standing topics in the physics of glass-forming liquids.[34,35,36,37] Aside of the translational and rotational motions of tracer molecules,[38,39] the decoupling of translational and rotational self diffusion in pure supercooled liquids was studied extensively, employing various experimental techniques[34,37] and molecular dynamics simulations.[36,40] Again different explanations were proposed, e.g., considering heterogeneity and dynamic correlation length scales,[36,38,41] a fluidized domain model,[35] or the mode coupling theory.[34,40] It should be noted that, in contrast to one-component supercooled liquids, the translation-rotation coupling or decoupling in the present case of dielectrically investigated DESs mainly refers to relations between the dynamics of different particles, namely the HBD molecules (rotation only) and the ions (translation only for the spherical Cl$^-$ ions and both translation and rotation for the charged and dipolar choline$^+$ ions).

It is obvious that a comparison of the temperature-dependent reorientational relaxation time $\tau_\varepsilon(T)$ and of the dc conductivity $\sigma_{dc}(T)$ with the viscosity $\eta(T)$ should help clarifying the mechanism of the rotation-translation coupling found in Ref. 13. Unfortunately, for the three considered systems glyceline, ethaline, and reline, such data are only available in a limited temperature range around room temperature, not allowing for any significant conclusions.[13] Therefore, in the present work, we investigate the viscosity of these DESs in a broad temperature range covering the high-viscosity supercooled regime, close to the glass temperature, and extending well into the liquid region. Moreover, via shear rheological spectroscopy we determine their frequency-dependent mechanical properties. The latter experiments reveal information on the viscous flow and thus on the translational dynamics of all constituents of the DES, in contrast to the conductivity which only mirrors ionic translations. According to the Maxwell relation $\tau = G_\infty/\eta$, where $G_\infty$ is the essentially temperature-independent high-frequency shear modulus, the structural relaxation times $\tau$ deduced from shear rheological spectroscopy ideally should be proportional to the viscosity $\eta$. As viscous flow has to involve translational motions, $\tau$ should reflect also translational dynamics. It can be compared to the reorientational relaxation times obtained from dielectric spectroscopy thus providing information on the rotation-translation coupling.

Notably, the dielectric data published in Ref. 13 for ethaline and reline were described by a Cole-Davidson function augmented by a dc-conductivity term only. In other words, it was not necessary to assume the presence of a significant ac-conductivity. This latter phenomenon is typically due to hopping charge transport, which often shows up in the complex conductivity $\sigma^* = \sigma' + i\sigma''$ and permittivity $\varepsilon^* = \varepsilon' - i\varepsilon''$ of ionic conductors. Such contributions are predicted by various models treating hopping conductivity of localized charge carriers (including ions)[42,43,44,45] as, e.g., the random free-energy barrier hopping model (RBM).[46,47] One interpretation of this finding is that in ethaline and reline the spectral features arising from dipolar reorientations completely superimpose any possible hopping charge-transport contributions to the dielectric spectra that may be present, except for the dc conductivity. (The latter, due to the common relation $\varepsilon'' \propto \sigma'/\nu$, leads to the usual $1/\nu$ divergence in $\varepsilon''(\nu)$. Similar behavior was also reported for ionic liquids.[24] However, in an earlier work, some of the present authors showed that the RBM can be applied to the rheological response of supercooled liquids as well.[48] More recently, it was demonstrated how the single-particle mean-square displacement can be derived from complex viscosity spectra.[49] Within this approach the rheological response should essentially reflect translational dynamics. In the same vein the spectral behavior predicted by the RBM to arise from hopping mass transport should be un-obscured by reorientational dynamics in these experiments. In the present work, we will check this notion for the investigated DESs.

## II. EXPERIMENTAL DETAILS

To prepare the DESs glyceline, ethaline, and reline (choline chloride mixed with glycerol, ethylene glycol, or urea, respectively; all with 1:2 molar ratio), appropriate amounts of the two constituents were mixed for 24 hours at 350 K inside a glass tube under argon atmosphere, using a magnetic stirrer. The resulting product was a colorless, clear liquid without any residuals. The water content of all samples was tested by coulometric Karl-Fischer titration and determined to be 0.18, 0.28, and 0.16 wt% for glyceline, ethaline, and reline, respectively. To prevent water uptake, during the measurements the samples were kept in dry nitrogen or helium atmosphere. For a description of the dielectric measurement techniques, the reader is referred to Ref. 13.

The mechanical investigations were performed using a parallel-plate configuration by means of an MCR 502 rheometer in combination with an EVU20 temperature control unit and a CTD450 oven from Anton-Paar. Using 4 and 12 mm geometries, the oscillatory shear measurements covered a frequency range from 0.001 to 100 Hz. The amplitude of the shear strain was adjusted to be in the linear-response regime of the measured stress. In order to determine the viscosity 12, 25, and 50 mm diameter plates were continuously rotated with shear rates between 0.01 and 100 1/s. Each measuring temperature was stabilized to within 0.2 K. Prior to carrying out the measurements, all samples were stored at 100 °C for more than 12 hours in order to reduce potentially present water traces.

## III. RESULTS AND DISCUSSION

### A. Dielectric spectra

As an example for the dielectric response of the investigated DESs, Fig. 1 shows spectra of the dielectric constant $\varepsilon'$, dielectric loss $\varepsilon''$, and the real part of the conductivity $\sigma'$ for glyceline.[50] The corresponding results for ethaline and reline were treated in detail in Ref. 13. Therefore, in the following we only briefly discuss the dielectric data, which are shown here to facilitate the direct comparison with the mechanical spectra. The $\varepsilon'$ spectra [Fig. 1(a)] are dominated by a strong increase towards low frequencies, showing up especially at the higher temperatures, which is of non-intrinsic nature and due to electrode polarization.[51] At sufficiently high frequencies or low temperatures, a clear step-like decrease in $\varepsilon'(\nu)$ (see inset for a zoomed view) evidences a relaxational process. With decreasing temperature, it strongly shifts to lower frequencies, reflecting the glassy freezing of this dynamics. Two third of this DES consists of glycerol, which exhibits a well-pronounced dipolar relaxation, and, especially at high temperatures, the relaxation steps in Fig. 1(a) occur at similar frequencies as for pure glycerol.[52] This is also revealed by a comparison of the relaxation times as given in Fig. 4 of Ref. 13.

Therefore, for high temperatures it is suggestive to assign the detected relaxation to the main intrinsic reorientational motions (commonly termed $\alpha$ relaxation[14,15]) in glyceline.[13,50] The increasing deviations of the relaxation times of glyceline and glycerol occurring upon cooling, as reported in Ref. 13, may be ascribed to the increasing relevance of intermolecular hydrogen-bond interactions at low temperatures in pure glycerol, which are partly broken up by the admixed ions in glyceline.

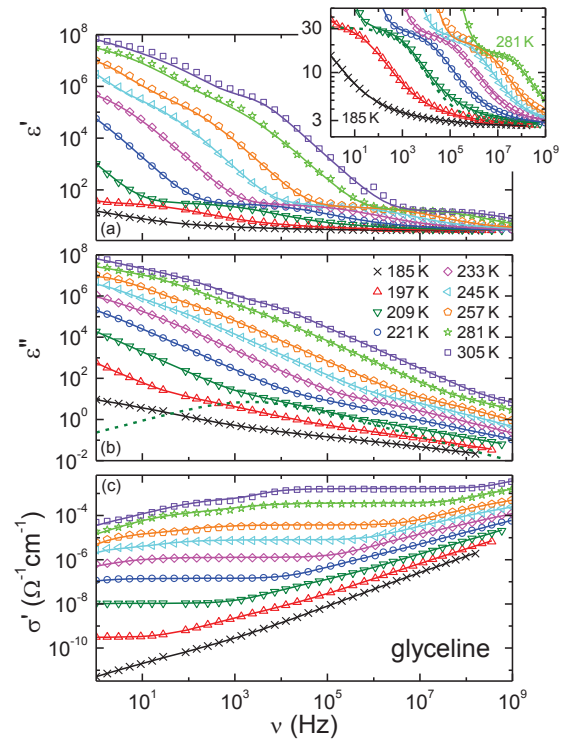

**FIG. 1.** Spectra of the dielectric constant $\varepsilon'$ (a), the dielectric loss $\varepsilon''$ (b), and the real part of the conductivity $\sigma'$ (c), measured at different temperatures for glyceline. The inset shows a zoomed view of $\varepsilon'(\nu)$ in the regime of the intrinsic relaxation. The solid lines in (a) and (b) are fits with an equivalent circuit, including up to two distributed RC circuits to account for the blocking electrodes, up to two intrinsic relaxational processes ($\alpha$ and $\beta$), and a dc-conductivity contribution (see text for details). $\varepsilon'(\nu)$ and $\varepsilon''(\nu)$ were simultaneously fitted. The fit lines in (c) were calculated using $\sigma' = \varepsilon'' \varepsilon_0 \omega$. The dashed lines in (b) and in the inset indicate the contribution of the $\alpha$ relaxation for 209 K.

The corresponding peaks in $\varepsilon''(\nu)$, expected to accompany the $\varepsilon'$ steps, are only visible as weakly-pronounced shoulders because their low-frequency flanks are superimposed by the conductivity contribution, $\varepsilon_{dc}'' \propto \sigma_{dc}/\nu$ [Fig. 1(b)]. The "dc" regime of conductivity is more directly evidenced by the frequency-independent regions

6The Journal of Chemical Physics

ACCEPTED MANUSCRIPT

This is the author's peer reviewed, accepted manuscript. However, the online version of record will be different from this version once it has been copyedited and typeset.
PLEASE CITE THIS ARTICLE AS DOI:10.1063/5.0045448

AIP Publishing

revealed in $\sigma'(\nu)$ [Fig. 1(c)]. The conductivity increase observed at frequencies beyond the dc region corresponds to the relaxational contribution in $\varepsilon''(\nu)$ (since $\sigma' \propto \varepsilon''\nu$). For the highest temperatures, at low frequencies $\sigma'(\nu)$ exhibits a decrease with decreasing frequency due to electrode blocking effects.[51] It corresponds to the flattening of the $\varepsilon''(\nu)$ curves, preceding the dc-related $1/\nu$ region as the frequency is increased.

For the quantitative description of the spectra two fitting approaches have been employed. One of them leads to the lines in Fig. 1 which are fits that formally account for the electrode polarization effects by (distributed) RC circuits[51] and describe the intrinsic relaxations by empirical Cole-Cole (CC) functions.[53] As an example, the dashed lines in the inset of Fig. 1 and in frame (b) illustrate the contribution of the $\alpha$ relaxation to the overall fit curve of the 209 K spectrum. The deviations at low frequencies are due to electrode polarization (for $\varepsilon'$) or the dc conductivity (for $\varepsilon''$). Within this approach the contributions emerging at high frequencies highlight the presence of a secondary relaxation or an excess wing as found in glycerol,[52] in the present approach mimicked by an additional CC function. Note that secondary relaxation processes with distinguishable dielectric loss peak maxima have been found for other DESs.[12,13] However, a detailed investigation of this spectral feature, requiring measurements at lower temperatures and/or higher frequencies, is out of the scope of the present work.

In most small-molecule glass formers, including glycerol, the $\alpha$ relaxation is described by a Cole-Davidson (CD)[54] function rather than a low-frequency-broadened CC function. Therefore, as a second fitting approach we employed a superposition of a CD function and a low-frequency ac conductivity contribution. For convenience, the latter is here parameterized using the 2008 version of the RBM.[47] The quality of the corresponding parameterization of the data is comparable to that of the former fitting approach, see Fig. 2.

In Ref. 13 it was shown that the dielectric response of ethaline cannot be fitted by the RBM alone,[46] while for several other glass formers this approach was found applicable.[55] As discussed in section I, the RBM was not conceived to describe spectra of materials with pronounced relaxation shoulders on their $\sigma'(\nu)$ responses, arising from reorientational motions. Showing the 221 K spectra as an example, Fig. 2, including a fit of the curves using only the RBM (dash-dotted lines), demonstrates the appearance of such a shoulder for glyceline. Clearly, assuming only ionic contributions, this model alone is not able to provide a good description of the dielectric relaxation data of glyceline, in harmony with previous findings for ethaline.[13] While the fits assuming a dipolar relaxation only (solid blue lines in Fig. 2) are able to match well the experimental spectra, they however, assume a CC-like spectral shape for the main relaxation process and thus a shape which is not really typical for the reorientational response of small-molecule glass forming liquids.

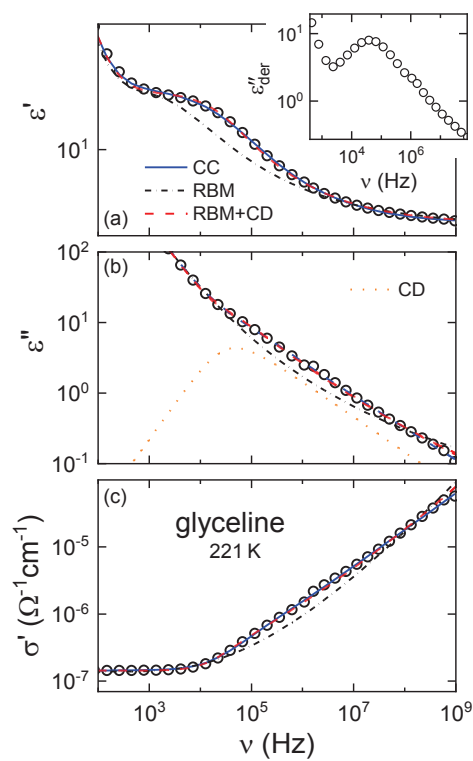

**FIG. 2.** Spectra of $\varepsilon'$ (a), $\varepsilon''$ (b), and $\sigma'$ (c) of glyceline at 221 K. The solid blue lines show the same fit curves as in Fig. 1, assuming a reorientational relaxation process described by a CC function for the intrinsic sample response. The dashed red line represents an alternative fit assuming an ac (relaxational) ionic contribution accounted for by the RBM[47] in addition to a CD-like dipolar contribution. The latter is indicated by the dotted orange line in frame (b). The dash-dotted lines demonstrate that the solution of the RBM alone is not able to describe the overall relaxational contributions well. In all cases, the non-intrinsic electrode-polarization contributions were modeled by a distributed RC circuit.[51] The inset shows an estimated loss spectrum as obtained by applying the derivative analysis of Ref. 62 to the $\varepsilon'$ spectrum.

Nevertheless, one should be aware that in some cases the RBM alone was successfully applied to systems featuring dipolar reorientations, including several ionic liquids and a few DESs.[12,49,56,57,58,59,60] It is clear that in such materials, ionic hopping – as treated by the RBM – and dipolar reorientations in principle could occur simultaneously and it may well be possible that in some cases the hopping contribution dominates (see also discussion of Fig. 5 in Ref. 55). Interestingly, in Ref. 58 spectra of an ionic liquid were successfully fitted assuming a superposition of the RBM prediction, in its 1988 version,[46] and a slower relaxational process. As shown in Ref. 52, the 2008 version of the RBM is, however, able to describe the data of ILs without the need to take such slow contributions into account. Interestingly, in Ref. 24, some of the present authors reported that the dielectric spectra of 12 ionic liquids could be well described by a sum of a CD function and a dc conductivity contribution. This observation has been interpreted in terms of dominant reorientational relaxations, without invoking any additional ac contributions arising from hopping charge transport. For one of these systems, it was explicitly shown that the RBM alone (in both variants)[46,47] cannot fully account for the experimental data. Overall, the application of the RBM seems to be straightforward for ionic conductors without dipolar degrees of freedom like the often-investigated ionic melt $[Ca(NO_3)_2]_{0.4}[KNO_3]_{0.6}$ (Ref. 61), but some care is necessary for materials involving dipolar reorientations.

To corroborate the interpretation of the observed shoulders in the loss and conductivity spectra in terms of partly hidden relaxation peaks [e.g., dashed lines in Figs. 1(b) and 2(b)], we have performed a derivative analysis of $\varepsilon'(\nu)$ as proposed in Ref. 62. Plotting the quantity $\varepsilon''_{der} = -\pi/2 \times \partial\varepsilon'/\partial\ln(\omega)$ should provide an estimate of the loss spectrum without any dc-conductivity contributions. In the inset of Fig. 2 we show an example for the 221 K data. Indeed, it leads to a clearly pronounced peak whose frequency position and amplitude reasonably agrees with the $\varepsilon''$ peak as resulting from our fit analysis [dashed line in Fig. 2(b)]. The deviations at the low-frequency flank of $\varepsilon''_{der}(\nu)$ are due to electrode polarization, affecting the $\varepsilon'$ spectra at low frequencies as discussed above.

Dielectric data for ionically conducting materials are often presented in terms of the complex dielectric modulus, $M^* = M' + i M'' = 1/\varepsilon^*$. Within the modulus formalism,[63] the peaks usually arising in $M''(\nu)$ are ascribed to a so-called conductivity relaxation and the relaxation times derived from them are assumed to provide a measure for the ionic mobility. It should be noted, however, that the use and interpretation of the modulus formalism is controversially discussed.[64,65,66,67] Figure 3 shows $M''$ spectra of glyceline for several temperatures revealing a clear two-contributions structure. It is well known that relaxation processes arising from dipolar reorientations also lead to a peak in $M''(\nu)$, with a peak position that is shifted to higher frequencies as compared to the $\varepsilon''$ spectra.[68,69,70] The high-frequency peaks in Fig. 3 reflect this effect. The low-frequency shoulders in the spectra of Fig. 3, then may be ascribed to the conductivity relaxation. However, one should be aware that the original modulus formalism was developed for "ionic conductors which contain no permanent molecular dipoles" like $[Ca(NO_3)_2]_{0.4}[KNO_3]_{0.6}$ or silicate glasses.[63] In these cases, the modulus peak usually occurs close to the crossover frequency between the dc plateau and a sublinear ac-conductivity increase in $\sigma'(\nu)$. The latter is considered typical for hopping charge transport[42,43,44,45,46] and the crossover frequency is often assumed to characterize the mobility of the ions.[45,46,71] In glyceline, the low-frequency modulus peak also occurs at the crossover of $\sigma'(\nu)$ from dc to a frequency-dependent region [cf. Figs. 1(c) and 3]. However, in view of the two fitting approaches discussed above, it is clear that the frequency dependence causing this crossover at least partly is governed by reorientational motions [cf., e.g., Fig. 2(c) and dotted line in Fig. 2(b)]. Thus, it is unclear whether or not the low-frequency modulus shoulder in Fig. 3 should be taken to provide information on the ionic dynamics. Thus, in the further course of this work, we refrain from evaluating the modulus data of the investigated DESs in detail. In our view, the most straightforward way to characterize the ion dynamics in such samples is by evaluating the dc conductivity, which can be unequivocally deduced from the $\sigma'(\nu)$ plateaus [Fig. 1(c)].

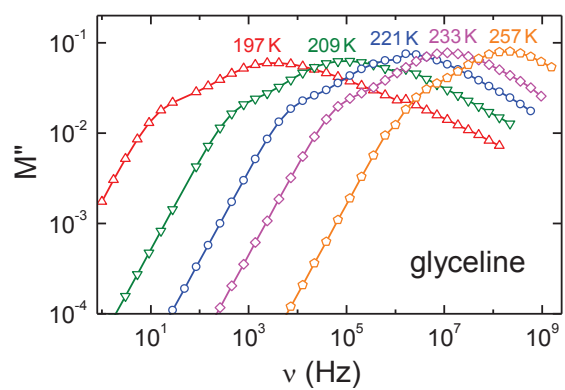

**FIG. 3.** Spectra of the imaginary part of the dielectric modulus $M''$ of glyceline, measured at various temperatures. The lines are guides for the eye.

### B. Rheological spectra

Results from rheological spectroscopy are often presented as spectra of the complex shear modulus, $G^* = G' + iG''$. As an example, Fig. 4(a) shows $G''(\nu)$ for glyceline at different temperatures. Clear peaks are revealed, strongly shifting to lower frequencies with decreasing temperature. This marks the slowing down of the structural shear relaxation in this DES when approaching the glass transition, just as found for the relaxation dynamics revealed by the dielectric spectra (Fig. 1). When comparing the mechanical and dielectric spectra, one should be aware that the data plotted in Fig. 1 are related to the dielectric susceptibility ($\chi^* = \varepsilon^* - 1$). In general, data from different spectroscopies should be compared in a consistent way, i.e., susceptibility with susceptibility in the present case.[72] Therefore, in Figs. 4(b) and (c), we plot the real and imaginary part of the mechanical compliance, $J^* = J' - iJ'' = 1/G^*$. As discussed, e.g., in Ref. 48, these quantities can be considered as the mechanical analogues of the dielectric quantities $\varepsilon'$ and $\varepsilon''$. Indeed, their spectral shapes qualitatively resemble the intrinsic contributions to the $\varepsilon'$ and $\varepsilon''$ spectra in Figs. 1(b) and (c). The steps observed in $J'(\nu)$ here indicate the structural relaxation process. The more limited frequency range of rheological measurements, compared to the dielectric ones, does not allow for the detection of the complete $J'(\nu)$ step for most temperatures. In principle, the relaxation should lead to a peak in $J''(\nu)$. However, similar to the dc conductivity in $\varepsilon''(\nu)$ [Fig. 1(b)], here the steady-state viscous flow leads to a $1/\nu$ divergence in $J''(\nu)$. This contribution in turn leads to a low-frequency plateau in the real part of the fluidity $F'(\nu) = 2\pi \nu J''(\nu)$, which can be regarded as the mechanical analogue of the conductivity $\sigma'$.[48,73] The "dc" fluidity $F_0$ corresponds to the inverse (steady-state) viscosity, $F_0 = 1/\eta$.

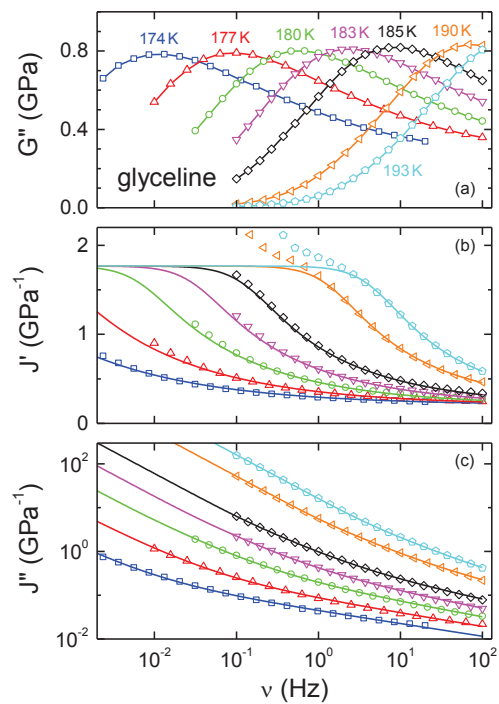

**FIG. 4.** Spectra of the imaginary part of the shear modulus (a) and of the real (b) and imaginary part (c) of the mechanical compliance, measured at various temperatures for glyceline. The lines in (a) are guides for the eye. The lines in (b) and (c) are fits with Eq. (1).

We found that the data of Fig. 4 can be well fitted using the phenomenological CD function,[54] adopted for the mechanical compliance and augmented by a term accounting for the steady-state flow:[55]

$$J^* = J_\infty + \frac{\Delta J}{\left(1 + i\omega\tau_{J0}\right)^\beta} + i\frac{1}{\omega\eta} \qquad (1)$$

Here, $J_\infty$ is the limiting high-frequency compliance, $\Delta J = J_0 - J_\infty$ the compliance relaxation strength with $J_0$ the so-called recoverable compliance, and $\beta$ is the width parameter. The fits were simultaneously performed for $J'$ and $J''$. However, due to the incomplete detection of the relaxation steps in $J'$, at most temperatures the parameters $J_\infty$ and $\Delta J$ had to be fixed to temperature-independent values to enable a reasonable determination of the other parameters. The fits then revealed a continuously increasing average relaxation time $\langle\tau_J\rangle$ (for the CD function, $\langle\tau_J\rangle = \beta\tau_{J0}$)[74] and viscosity with decreasing temperature, which will be discussed in section IIIC in detail. Upon cooling, $\beta(T)$ from the fits decreases from 0.4 to 0.29.

An alternative way to evaluate such data is the construction of a so-called master curve by horizontally and vertically shifting the (logarithmic) compliance spectra, that were recorded at different temperature, onto a reference spectrum. If time-temperature superposition holds, i.e., if the shape of the spectra does not change with temperature, a single master curve can be obtained this way which then covers a considerably extended frequency range. For the present glyceline data, Fig. 5 shows the master curves for $J'$, $J''$, and $F'$ using the spectrum at 190 K as reference. A good match could be achieved, indicating that time-temperature superposition indeed is valid for these data, at least in the somewhat limited frequency range in which the mechanical spectra were taken. Notably, no vertical shift was necessary to arrive at these master curves. This justifies the assumption of constant $J_\infty$ and $\Delta J$ made for the fits of the spectra, see Fig. 4.

As shown by the solid lines in Fig. 5 the master curves can reasonably be fitted by the CD function, Eq. (1), leading to $\langle\tau_J\rangle \approx 73$ ms and $\beta \approx 0.34$. Only at the highest frequencies in $J''$ (and, thus, in $F'$) some deviations show up, probably reflecting the mentioned reduction of the width parameter $\beta$ for the low-temperature spectra. The dashed lines in Fig. 5 provide alternative fits of the master curves using the RBM prediction,[47] modified for the description of mechanical compliance data.[48,75] In contrast to the dielectric data (Fig. 2), for the mechanical spectra obviously the RBM alone (i.e., without adding other contributions) is able to provide fits of comparable quality as the CD approach. As discussed in section I, this different behavior for the two spectroscopies can be rationalized if one considers that the shear mechanical data exclusively reflect translational dynamics[49] as modeled by the RBM,[48] which is not the case for the dielectric spectra. For the characteristic time parameter of the RBM, the fits lead to $\tau_{RBM} \approx 53$ ms, of comparable magnitude as $\langle\tau_J\rangle$. Finally, from the shift factors used to generate the master curves and from the relaxation time obtained from its fit, temperature-dependent values for $\langle \tau_J \rangle$ and $\tau_{RBM}$ can be deduced. They agree well with $\langle \tau_J \rangle(T)$ resulting from the analysis of the spectra shown in Fig. 4.

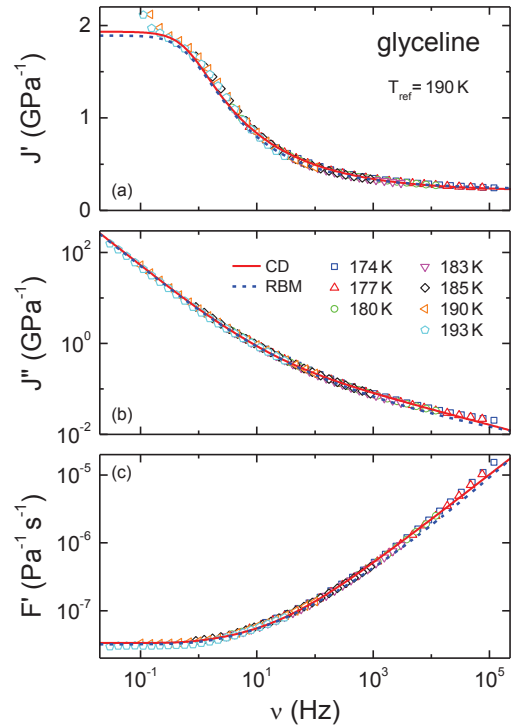

**FIG. 5.** Master curves of glyceline for the real (a) and imaginary part (b) of the mechanical compliance and the real part of the fluidity (c). The spectra measured at different temperatures were horizontally shifted to the respective reference curve at 190 K. The solid and dashed lines show the results of simultaneous fits of the $J'$ and $J''$ master curves by Eq. (1) or by the RBM formula adapted for rheological data,[48] respectively.

For reline, the compliance and fluidity spectra look qualitatively similar to those for glyceline.[76] Figure 6 shows the resulting master curves for the complex shear compliance of reline. While for $J''(\nu)$, a nearly perfect master curve could be obtained, in $J'(\nu)$ significant deviations show up at low frequencies. We ascribe these to spurious contributions in $J'(\nu)$ due to the low-torque limitations of the rheometer.[48,49,55,61] To a lesser extent they also affect the glyceline spectra [cf. the two highest temperatures in Fig. 4(b)]. Again, both, the CD function, Eq. 1, and the modified RBM[48,] formula can be employed to fit the compliance master curves of reline. We obtain $\langle \tau_J \rangle \approx 46$ ms and $\beta \approx 0.32$ from the CD fit and $\tau_{RBM} \approx 30$ ms from the RBM fit.

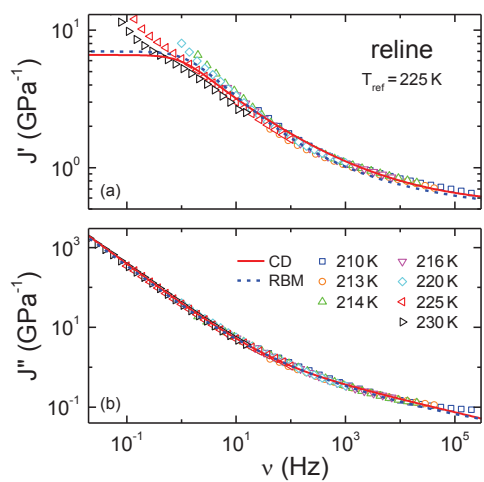

**FIG. 6.** Master curves of reline for the real (a) and imaginary part (b) of the mechanical compliance. The spectra measured at different temperatures were horizontally shifted with respect to the reference curve at 225 K. The solid and dashed lines show the results of simultaneous fits of the $J'$ and $J''$ master curves by Eq. (1) or by the RBM formula adapted for rheological data,[48] respectively.

We found ethaline to exhibit a higher tendency to crystallize than glyceline and reline and, thus, the dielectric measurements were done with somewhat increased cooling rates (1 K/min instead of 0.4 K/min) to avoid crystallization. However, this enhanced crystallization tendency hampers dynamic rheology measurements, because they require relatively long sweeping times. Hence, for ethaline we were not able to collect meaningful mechanical-spectroscopy data. At least, steady-state viscosity data could be determined at several temperatures using the rotation and oscillation modes of the spectrometer, which will be discussed below.

### C. Coupling of reorientational and translational dynamics

Figure 7 provides an Arrhenius plot of the main dynamical quantities of glyceline, obtained from the analysis of the dielectric and rheological measurements presented in the previous section. The crosses show the dielectric relaxation times, characterizing the reorientational dynamics according to the CC approach discussed above. The plusses represent the structural relaxation times from the mechanical compliance spectra, mirroring translational motions, in principle of all constituents of the material. Essentially the same dynamics also governs the viscosity shown by the triangles (the latter also include data from literature around room temperature).[77] Finally, the circles depict the dc resistivity $\rho_{dc} = 1/\sigma_{dc}$ derived from the dielectric measurements that allow to monitor the translational ion motions.[78] As already reported for $\rho_{dc}(T)$ and $\langle\tau_\varepsilon\rangle(T)$,[13] strong deviations from simple thermally activated Arrhenius behavior also show up for $\eta(T)$ and $\langle\tau_J\rangle(T)$. As explicitly demonstrated for $\rho_{dc}(T)$ in Fig. 7 (solid line), all data sets can be well described by the Vogel-Fulcher-Tammann (VFT) law,[79,80,81] typical for glassy dynamics, adapted to the respective quantity $q$:

$$q = q_0 \, exp\left[\frac{DT_{VF}}{T - T_{VF}}\right] \quad (2)$$

Here $q_0$ is a pre-exponential factor, $D$ is the so-called strength parameter quantifying the deviation from Arrhenius behavior[82] [recovered from eq. (2) for $T_{VF} \to 0$], and $T_{VF}$ is the so-called Vogel-Fulcher temperature, where $q$ would diverge.

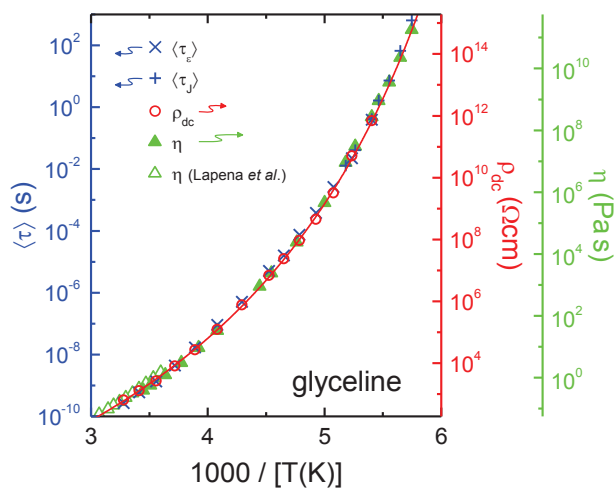

**FIG. 7.** Arrhenius plot of the dynamical quantities characterizing the translational and reorientational dynamics of glyceline as deduced from dielectric and rheological measurements. Shown are the average dielectric relaxation time $\langle\tau_\varepsilon\rangle$ determined via the CC approach (crosses; left scale), the structural relaxation time from compliance measurements $\langle\tau_J\rangle$ (plusses; left scale), the dc resistivity $\rho_{dc}$ (circles; first right scale), and the viscosity $\eta$ (closed triangles; second right scale). For the latter, additional data at high temperatures from literature[77] are included (open triangles). The three ordinates were adjusted to cover the same number of decades and their starting values were chosen to reach a good match of the different quantities. The line is a VFT fit of $\rho_{dc}$ ($\rho_{dc,0} = 1.2\times 10^{-2}$ Ωcm, $D = 16$, $T_{VF} = 123$ K).

To enable a direct comparison of the different quantities, in Fig. 7 the respective ordinates were adjusted to cover the same number of decades (13 to be specific). Consequently, we find that, by a proper choice of the ordinate starting-values, it is possible to achieve a good match of the temperature-dependent traces of all these different quantities within experimental uncertainty (about the size of the symbols). This proves a direct coupling of the different types of dynamics, i.e., ionic and molecular translation as well as dipolar reorientation, and implies the validity of the proportionalities suggested by the Stokes-Einstein, Debye-Stokes-Einstein, Nernst-Einstein, and Maxwell relations, mentioned in section I. Thus, in principle, one can understand the close correlation of the dipolar and ionic motions in glyceline, reported in Ref. 13, by the coupling of each of these dynamics to the viscosity, without invoking a more direct link via a revolving-door mechanism.

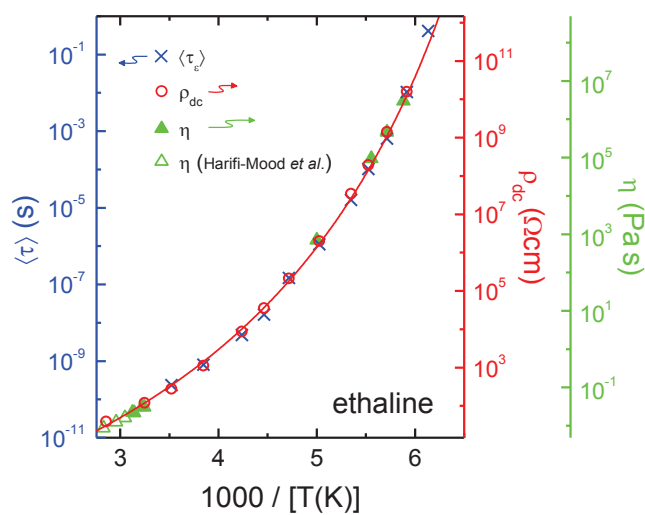

**FIG. 8.** Comparison of the dynamical quantities of ethaline (corresponding representation and same symbol meanings as in Fig. 7). The viscosity literature data (open triangles) are from Ref. 83. The line is a VFT fit of $\rho_{dc}$ ($\rho_{dc,0}$ = 5.3×10$^{-2}$ Ωcm, $D$ = 13.7, $T_{VF}$ = 111 K).

Figure 8 shows a comparison of the dynamical quantities for ethaline (representation analogous to that in Fig. 7). Owing to the mentioned crystallization tendency of this DES, structural relaxation times from shear compliance spectroscopy could not be determined here. Nevertheless, the available viscosity data (from the present work and Ref. 83), together with the dielectric results ($\langle \tau_\varepsilon \rangle$ and $\rho_{dc}$), are sufficient to infer a good coupling of all detected dipolar and translational motions in this material, too. Thus, for ethaline, the proportionality of $\langle \tau_\varepsilon \rangle$ and $\rho_{dc}$ may also be understood in terms of a coupling of both quantities to the viscosity, according to the Stokes-Einstein and Debye-Stokes-Einstein relations. Very recently, the rotational dynamics in ethaline at room temperature was studied using molecular dynamics simulations.[84] The main relaxation times deduced from the decay of the rotational correlation functions of choline and ethylene glycol (0.35 and 0.08 ns, respectively) are of a similar order of magnitude as $\langle\tau_\varepsilon\rangle \approx 0.11$ ns derived from our dielectric results. It should be noted that dielectric spectra of liquids with two dipolar components (as choline and ethylene glycol in the present case) often exhibit a single $\alpha$ relaxation process only.[85,86]

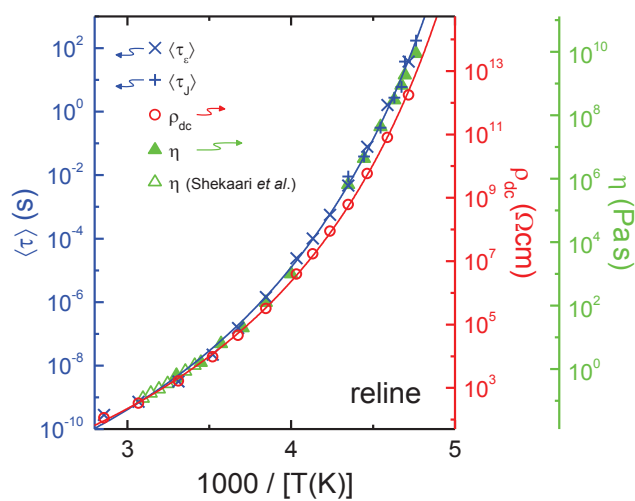

**FIG. 9.** Comparison of the dynamical quantities of reline (corresponding representation and same symbol meanings as in Fig. 7). The viscosity literature data (open triangles) are from Ref. 87. The lines are fits of $\langle\tau_\varepsilon\rangle$ ($\tau_0 = 4.0\times10^{-15}$ s, $D = 13$, $T_{VF} = 157$ K) and of $\rho_{dc}$ ($\rho_{dc,0} = 8.9\times10^{-3}$ $\Omega$cm, $D = 11$, $T_{VF} = 159$ K).

Finally, in Fig. 9 we provide a corresponding comparison plot for reline. Interestingly, here the situation is different. While the reorientational and structural relaxation times and the viscosity (including literature data from Ref. 87) reasonably match each other, the dc resistivity exhibits significant deviations. In this graph, we have adjusted the starting value of the $\rho_{dc}$ ordinate to arrive at a good match with the other quantities at the highest temperatures, because decoupling effects usually become most prominent at low temperatures. Then, at the lowest temperatures, the resistivity data deviate from the other quantities by about one decade. While the decoupling of dipolar and ionic dynamics of reline at low temperatures was already detected in Ref. 13, Fig. 9 reveals that only the ion dynamics deviates from the common temperature dependence, while all other quantities, including viscosity, structural, and reorientational relaxation are well coupled to each other. It seems that in reline the ions have higher mobility than expected, based on the viscosity or (within the revolving-door picture) based on the reorientational motions.

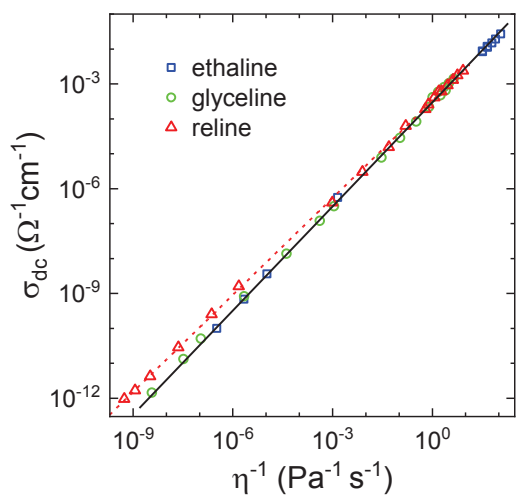

**FIG. 10.** Dc conductivity *vs.* inverse viscosity for the three investigated DESs, in a representation that is analogous to a Walden plot. The conductivities at the given temperatures of the viscosity measurements were determined by a B-spline interpolation of the $\rho_{dc}(1/T)$ curves, Figs. 7 - 9. The solid line with slope 1 indicates linear behavior, $\sigma_{dc} \propto \eta^{-1}$, for glyceline and ethaline. The dashed line is a linear fit of the reline data with slope 0.93, implying a fractional Walden law, $\sigma_{dc} \propto \eta^{-0.93}$.

In Fig. 10 the dc conductivity of the three investigated DESs is plotted *vs.* the inverse viscosity. Because of the identical ion concentration of the three DESs, Fig. 10 can be considered as analogous to a Walden plot,[88] where the molar conductivity $\Lambda$ ($\sigma_{dc}$ divided by the ion concentration) is plotted instead of $\sigma_{dc}$.[89] The data for glyceline and ethaline can be well described by a linear increase (solid line), evidencing the validity of the Walden rule,[90] $\Lambda \propto \eta^{-1}$. These two data sets agree within experimental uncertainty, i.e., the same viscosity allows for the same ionic mobility, despite the difference in the HBDs. It should be noted that in these two DESs the Walden rule works over a broad conductivity and viscosity range, down to low temperatures approaching $T_g$, which, to our knowledge, was never before tested for these or other DESs.

For reline, Fig. 10 reveals clear deviations from a linear behavior, and we find a power law, $\sigma_{dc} \propto \eta^{-0.93}$, instead. This corresponds to a so-called fractional Walden rule[88] as was also reported for several ionic liquids.[91,92] For some DESs (e.g., Refs. 93,94), including reline,[95] Walden plots were previously reported in the literature. However, in general the investigated viscosity range (i.e., temperature range) was too limited to reliably check for deviations from the Walden rule. Again, Fig. 10 evidences that at high viscosities and low temperatures, the ions in reline exhibit higher mobility than expected for a conventional ionic conductor. Notably, similar effects (but much more pronounced) lead to the high conductivities of superionic glasses.[88]

## IV. SUMMARY AND CONCLUSIONS

In summary, we have performed rheological measurements of the typical DESs glyceline, ethaline, and reline, providing the basis for a thorough comparison of the obtained results with those from dielectric spectroscopy. Our shear experiments cover a broad temperature and dynamic range, extending from the low-viscosity liquid state down to temperatures close to the glass transition. As previously reported for ethaline and reline,[13] the dielectric spectra for glyceline were interpreted to involve dipolar reorientation processes. Furthermore, we found that the glyceline spectra could also not be explained by translational ionic charge transport alone. The mechanical compliance spectra reported in the present work could be described by a recently proposed adaptation of the RBM, suggestive of the presence of translational dynamics for rheological measurements[48] or alternatively and equally well by a CD function.

From the rheological measurements we have extracted information about the viscosity and structural relaxation time. Their temperature variation over up to 12 decades is clearly non-Arrhenius and approximately follows the VFT equation. Viscosity and structural relaxation time are compared to the dc conductivity, mirroring ionic translations only, and to the dielectric relaxation time characterizing reorientational processes.

For two of the presently studied DESs, glyceline and ethaline, we find a good coupling of all these quantities, as predicted by the Stokes-Einstein, Debye-Stokes-Einstein, and Nernst-Einstein relations and by the Walden rule. Thus, the rather naive picture of a sphere that translationally moves within a viscous medium or of an asymmetric particle that rotates within such a medium, leading to a complete coupling of both dynamics to the viscosity, seems to be valid in these DESs. Therefore, while a revolving-door scenario still may play some role in these materials, the previously reported close connection of their reorientational and ionic dynamics[13] can be also explained without invoking this mechanism.

In reline, a close coupling to the viscosity and structural relaxation was also found for the reorientational motions but, interestingly, not for the ionic mobility. Especially, the ionic dc conductivity follows a fractional Walden rule as previously revealed in some ionic liquids.[91,92] It seems that in this DES a charge-transport mechanism is active that enhances the ionic mobility beyond the above-mentioned naive picture. Notably, the technically relevant charge transport is the only dynamics that exhibits a decoupling in reline, which implies that the revolving-door mechanism cannot explain its non-canonical conductivity behavior. It should be noted that the hole theory, often employed to understand ionic charge transport in DESs on a microscopic level, predicts the validity of the Walden rule.[96] Interestingly, based on neutron diffraction it was suggested that the hole theory may not be applicable to reline,[97] consistent with the present finding of a violation of this rule.

Achieving an understanding of the special behavior of reline, compared to glyceline and ethaline, is complicated by the fact that different ion species can contribute to the charge transport in these systems and that complex aggregation between ions and the HBD molecules can arise.[84,97,98,99,100,101] For example, using charge decomposition analysis, Wagle et al.[101] reported that considerable charge transfer occurs from the ions to the HBDs in some choline-chloride-based DESs, including ethaline, but that this effect is only weak in reline. Based on nuclear magnetic resonance[99] and neutron scattering,[84] the HBDs in glyceline, ethaline, and reline were found to exhibit significantly faster translational and rotational diffusion than the choline ions. Therefore, in glyceline and ethaline a significant contribution from the HBDs to the charge transport can be assumed which, because of the reduced charge transfer, is not the case for reline.[101] Moreover, it was noted that in DESs strong hydrogen bonds often form between the anions and the HBDs, which should lead to the presence of complex [HBD]Cl$^{-1}$ anions within the choline-chloride-based DESs.[98,99,100] However, based on neutron scattering and atomistic modeling, for reline the formation of an even more complex supramolecular structure, comprising a choline and chloride ion and two urea molecules, was reported.[97] Finally, by molecular dynamics simulations, Zhang et al.[84] arrived at a significantly different Cl$^-$ solvation environment in ethaline compared to reline. All these findings make it appear reasonable that the charge transport in reline is rather special and occurs by a mechanism different from that active in glyceline and ethaline, as suggested by the present results. However, its microscopic details are still far from being understood and certainly more investigations on related DESs (e.g., maline, which reveals a slower diffusion of the HBD compared to choline[97]) would be helpful for a deeper understanding of the charge transport in this class of DESs.

**ACKNOWLEDGMENTS**

This work was supported by the Deutsche Forschungsgemeinschaft (grant No. LU 656/5-1 and partly by grant No. GA2680/1-1).

## DATA AVAILABILITY

The data that support the findings of this study are available from the corresponding author upon reasonable request.